\documentclass[12pt]{article}
\usepackage{amssymb}
\usepackage{bm}
\usepackage{graphicx}

\newcommand{\hf}{{\hat{\phi}}}
\newcommand{\hx}{\hat{x}}
\newcommand{\hk}{\hat{k}}
\newcommand{\hP}{\hat{\partial}}

\def\he{{\hat e}}
\usepackage{bm}
\newcommand{\be}{\begin{equation}}
\newcommand{\ee}{\end{equation}}
\newcommand{\bea}{\begin{eqnarray}}
\newcommand{\eea}{\end{eqnarray}}
\begin{document}

\title{ $\kappa$-Minkowski space, scalar field, and the issue of Lorentz invariance}

\author{Laurent Freidel\thanks{Perimeter Institute,
Waterloo, Canada, \tt{lfreidel@perimeterinstitute.ca}}\,\,  and Jerzy Kowalski-Glikman\thanks{Institute for Theoretical Physics, University of Wroclaw, Wroclaw, Poland, \tt{jkowalskiglikman@ift.uni.wroc.pl}} %
}\maketitle

\begin{abstract}
We describe $\kappa$-Minkowski space and its relation to group theory. The group theoretical picture makes it possible to  analyze the symmetries of this space. As an application of this analysis we analyze in detail  free field theory on $\kappa$-Minkowski space and the Noether charges associated with deformed spacetime symmetries.
\end{abstract}

\section{Introduction}

$\kappa$-Minkowski space \cite{kappaM1}, \cite{kappaM2} is a particular example of non-commutative space, in which positions $\hat x^\mu$ satisfy the algebra-like commutational relation between ``time'' and ``space''\footnote{We set the deformation scale $\kappa=1$ in what follows.}
\begin{equation}\label{1}
 [\hat x^0, \hat x^i] = i \hat x^i
\end{equation}
with all other commutators vanishing. Such space arouse first in the investigations of $\kappa$-Poincare algebra \cite{kappaM1}, \cite{kappaM2}. Later it has been related to Doubly Special Relativity (see \cite{Kowalski-Glikman:2004qa} for review and references) and it has been claimed that it has a quantum gravitational origin \cite{Amelino-Camelia:2003xp}, \cite{Freidel:2003sp}. If this claims are correct, $\kappa$-Minkowski space is to replace the standard Minkowski spacetime in description of ultra high energy processes, in the limit when (quantum) gravitational effects could be regarded as negligible.

Only recently however a theory of fields living on this space has started being analyzed in depth \cite{Agostini:2006nc}, \cite{Arzano:2007gr}, \cite{Freidel:2006gc}, \cite{Freidel:2007hk}. Thanks to the results reported in these papers we are now  not only understanding quite well the structure of $\kappa$-Minkowski space, and its relation to group theory, but also we understand free scalar field theory on this space, including the way how to construct conserved Noether charges associated with its symmetries.

In this paper we would like to describe this recent progress. Our goal is however not to repeat results of our recent paper \cite{Freidel:2007hk} but to explain what is its main message. Thus we spend some time discussing the structure of $\kappa$-Minkowski space. Then we formulate scalar field theory on this space, and after quoting results from this paper, we try to analyze their physical meaning.

\section{Group theory and deformed Poincar\`e symmetry of $\kappa$-Minkowski space}

Before starting our investigations, let us first introduce the notion of co-product, which is going to be crucial in what follows. Consider eq.\ (\ref{1}). As it stands it looks Lorentz non-covariant. But is it indeed? Lets see.

Assume that Lorentz generators, of rotation $M_i$ and boost $N_i$ act on positions in the standard way, as follows
$$
 M_{i}\triangleright \hat{x}_{0}=0,\;\;\; M_{i}\triangleright \hat{x}_{j}=i\epsilon_{ijk}\hat{x}_{k},
$$
 \begin{equation}\label{2}
 N_{i}\triangleright \hat{x}_{0}=i\hat{x}_{i},\;\;\; N_{i}\triangleright \hx_{j}=i\delta_{ij}\hx_{0}.
\end{equation}
This however does not say how the generators act on product of position. Usually one applies Leibniz rule, for example
$$
N_{i}\triangleright (\hat{x}_{0}\, \hat x_j )= (N_{i}\triangleright \hat{x}_{0})\, \hat x_j +  \hat{x}_{0}\, (N_{i}\triangleright\hat x_j)
$$
and then the left hand side of (\ref{1}) transforms differently than the right hand side. But Leibniz rule is not sacred, it can be replaced by something more general. Let us try the following rule
$$
N_{i}\triangleright (\hat{x}_{0}\, \hat x_j )= (N_{i}\triangleright \hat{x}_{0})\, \hat x_j +  \hat{x}_{0}\, (N_{i}\triangleright\hat x_j) + i \, (N_{i}\triangleright\hat x_j) = i \hx_i\, \hx_j + i \hx_0\, \hx_0 \delta_{ij} - \hx_0 \delta_{ij}
$$
$$
N_{i}\triangleright (\hat{x}_{j}\, \hat x_0 )= (N_{i}\triangleright \hat{x}_{j})\, \hat x_0 +  \hat{x}_{j}\, (N_{i}\triangleright\hat x_0)  =   i \hx_0\, \hx_0 \delta_{ij} + i \hx_j\, \hx_i
$$
Subtracting and noticing that $\hx_i\, \hx_j = \hx_j\, \hx_i$ we find that the action of boost on commutator equals $- \hx_0 \delta_{ij}$ which is exactly $iN_{i}\triangleright \hx_j$! Thus we saved covariance of the $\kappa$-Minkowski defining relation, eq.\ (\ref{1}). The price we had to pay was the deviation from Leibniz rule. In the theory of Hopf algebras such deviation is called coproduct, it says how to act with an algebra on a products of representations. In more abstract terms one the coproduct is defined as mapping from an algebra to tensor product of it
 $   \vartriangle: {\cal A} \rightarrow {\cal A}\otimes {\cal A}$;
we recover the standard Leibniz rule by taking trivial co-product: $\vartriangle({\cal A}) = \mathbf{1} \otimes {\cal A} + {\cal A} \otimes \mathbf{1}$. The rule of action of Lorentz generators on product of positions is a particular example of the nontrivial co-product structure of $\kappa$-Poincar\`e algebra, being the algebra of symmetries of $\kappa$-Minkowski space
 \begin{equation}\label{3}
\triangle (M_i)=M_i\otimes \mathbf{1}+\mathbf{1}\otimes M_i,\quad    \triangle (N_i)=N_i\otimes \mathbf{1} +e^{-{k_0}}\otimes N_i+\epsilon_{ijk}k_j\otimes M_k
\end{equation}
Notice that the coproduct of rotations, $M_i$ is trivial, and thus for them we have to do with the standard Leibniz action. In the formula above $k_\mu$ are some generators of translation which satisfy
\begin{equation}\label{4}
    k_\mu \triangleright \hx^\nu = i \delta^\nu_\mu
\end{equation}
Using this and (\ref{3}) one can easily reproduce the the formulas presented above and check that also the commutator $[\hx^i, \hx^j]=0$ transforms covariantly. The origin of formulas (\ref{3}) is not completely clear yet, but we will return to them in a moment.

Before doing so let us notice the important difference between the action of Lorentz generator on positions defined in (\ref{2}) and the commutator. Indeed the latter is defined to be (we consider boosts only because for rotation, as a result of trivial coproduct the result is the same as in the classical case)
$$
[N_i, \hx_0] \triangleright (\star) \equiv N_i \triangleright [\hx_0 \triangleright (\star)] - \hx_0 \triangleright [N_i \triangleright (\star)]
$$
where the  position acts by multiplication.
For example
$$
[N_i, \hx_0] \triangleright \hx_j = N_i \triangleright (\hx_0\hx_j) - i \hx_0 \hx_0 \delta_{ij} = i \hx_i\, \hx_j  - \hx_0 \delta_{ij} = i \hx_i \triangleright \hx_j +i N_i \triangleright \hx_j
$$
so that
$$
[N_i, \hx_0] = i \hx_i + i N_i
$$
Similarly one can derive the form of the commutator $[N_i, \hx_j]$.

Let us now turn to our main theme, which is group theory. It is obvious that the defining relation of $\kappa$-Minkowski space (\ref{1}) is a Lie algebra type one (contrary to the so-called canonical non-commutativity investigated in the context of string theory). It is surprising, a posteriori, that serious investigations of the group structure associated with it have begun only recently.

To start consider the following $5\times5$ matrix representation of the generators $x^\mu$
\begin{equation}\label{5}
\hat x^0 = -{i} \,\left(\begin{array}{ccc}
  0 & \mathbf{0} & 1 \\
  \mathbf{0} & \mathbf{0} & \mathbf{0} \\
  1 & \mathbf{0} & 0
\end{array}\right) \quad
\hat{\mathbf{x}} = {i} \,\left(\begin{array}{ccc}
  0 & {\bm{\epsilon}\,{}^T} &  0\\
  \bm{\epsilon} & \mathbf{0} & \bm{\epsilon} \\
  0 & -\bm{\epsilon}\,{}^T & 0
\end{array}\right),
\end{equation}
where $\bm{\epsilon}$ is a three dimensional vector with a single unit entry. Notice now that $x^0$ generates abelian subalgebra while the generators corresponding to spacial positions $\mathbf{x}$ are nilpotent $\mathbf{x}^2=0$. For this reason mathematicians denote such algebra $\mathfrak{a}\mathfrak{n}(3)$, and the corresponding group $\mathfrak{A}\mathfrak{N}(3)$. Such algebras and groups arise so-called Iwasawa decomposition. Following \cite{Freidel:2007hk} we will use the name Borel algebra and group.

A Borel group element can be written as
\begin{equation}\label{6}
   \he_k \equiv e^{ik_i \hat x^i} e^{ik_0 \hat x^0}
\end{equation}
(If we interpret $k$ as momentum this can be interpreted as a ``plane wave on $\kappa$-Minkowski space'' \cite{Amelino-Camelia:1999pm}.) The first natural question is what is the group manifold of Borel group. To answer it let us consider the matrix representation of (\ref{6})
\begin{equation}\label{7}
    \he_k = K_A{}^B= \left(\begin{array}{ccc}
  \bar P_4 & -\mathbf{P}e^{-k_0} & P_0 \\
  -\mathbf{P} & \mathbf{1} & -\mathbf{P} \\
  \bar P_0 & \mathbf{P}e^{-k_0} &  P_4
\end{array}\right)
\end{equation}
where $(P_0, \mathbf{P}, P_4)$ are given by\footnote{$\bar P$ are defined similarly, and the exact expressions can be found in \cite{Freidel:2007hk}.}
\begin{eqnarray}
 {P_0}(k_0, \mathbf{k}) &=&  \sinh
{{k_0}} + \frac{\mathbf{k}^2}{2}\,
e^{  {k_0}}, \nonumber\\
 P_i(k_0, \mathbf{k}) &=&   k_i \, e^{  {k_0}}, \nonumber\\
 {P_4}(k_0, \mathbf{k}) &=&  \cosh
{{k_0}} - \frac{\mathbf{k}^2}{2}\, e^{  {k_0}}.
\label{8}
\end{eqnarray}
It is easy to check that they satisfy the conditions
\begin{equation}\label{9}
   -P_0^2 + \mathbf{P}^2 + P_4^2 =1, \quad P_0 + P_4 \geq0
\end{equation}
Now if we act on a unit vector $(0,0,0,0,1)^T$ with the matrix $K_A{}^B$ we obtain points in $5d$ space with coordinates (\ref{8}), i.e., all points satisfying (\ref{9}). But this is nothing but a half of de Sitter space, see Fig.\ 1. Thus the momenta labeling of plane waves belong not to the flat space as usual, but to (a submanifold of) curved de Sitter space. In the construction of field theory we will have to take the curvature and global structure of the manifold (\ref{9}) into account.

\begin{figure}
\begin{center}
\includegraphics[width=.6\textwidth]{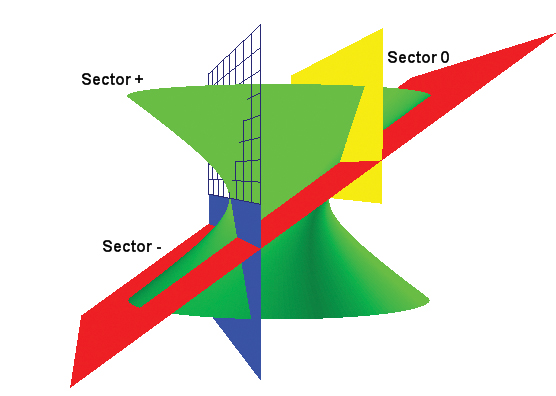}
\end{center}
\caption{The momentum space is the portion of De Sitter space above the plane $P_+=0$ where $P_{0}$ is the vertical axis. The mass shell is given by the intersection of this portion of de Sitter space with the vertical planes $P_{4}=\pm cste$. This mass shell naturally decomposes into three sectors indicated: $+$ with positive energy and $P_4>0$, $-$ with negative energy and $P_4>0$, and $0$ with  positive energy and $P_4<0$. Notice that in the limit $\kappa\rightarrow\infty$ the second sector becomes unbounded, while the third sector disappears.}
\end{figure}


Let us investigate the structure of the Borel group in more details. Consider composition of elements of the form (\ref{6}).
\begin{equation}\label{10}
    \he_{kl} \equiv \he_k \he_l = e^{i\hat x^i(k_i + e^{-k_0}l_i)} e^{i \hat x^0(k_0 + l_0)}
\end{equation}
The composition of  group elements (``plane waves'') can be equivalently described in terms of a non trivial  Hopf algebra structure for the momentum $k$, the co-product.
 Since $k$ can be regarded as a function on Borel group, one can associate with it the non commutative co-product dual to the group multiplication, which turns out to be
 \begin{equation}\label{11}
   \Delta(k_i) = k_i \otimes 1 + e^{-k_0} \otimes k_i, \quad \Delta(k_0) = k_0 \otimes 1 + 1 \otimes k_0
\end{equation}
Similarly the conjugate of a plane wave
\begin{equation}\label{12}
   (\he_k)^\dag = e^{-i  k_0 \hat{x}^0} e^{-i   k_i \hat{x}^i} = e^{-i \hat  (e^{k_0}k_i)\hat{x}^i}e^{-i  k_0 \hat{x}^0} = \he_{S(k)}
\end{equation}
gives the antipode
\begin{equation}\label{13}
   S(k_i) = - e^{k_0}k_i, \quad S(k_0) = - k_0.
\end{equation}
which is another object known in the Hopf algebras theory. We see therefore that group theory of Borel group is on one to one correspondence with the Hopf algebra structure of the space of momenta.

At the beginning of this section we discussed Lorentz transformations of positions. Now we can deduce how these transformations act on momenta. To this end let us act with such transformation on a plane wave
\bea\label{14}
 N_i\vartriangleright\hat{e}_k&=& i\, \left( {1\over 2} \left(1 -e^{-2{{k}_{0}}}\right) + {1\over 2} {\mathbf{k}}\,{}^{ 2}\, \right)  :
 \hat{x}_{i}\hat{e}_{k}:  - i\, {k}_{i} : \left({\mathbf{k}}{\mathbf{\hat{x}}} +\hat{x}_{0}\right)  \hat{e}_k: \\
 M_i\vartriangleright\hat{e}_k &=& i\epsilon^{ijk} k_{j}: \hat{x}_{k} \hat{e}_{k}:
 \eea
 where $:f(\hx):$ means ordered function with all $\hx_0$ shifted to
the right.

By moving $\hat{x}^{\mu}$ out of the normal ordering (\ref{14}) we can simplify the action of
Lorentz transformations which then read
 \bea\label{15}
  N_i\vartriangleright\hat{e}_k&=&
 i\left(\hat{x}_i P_0(k)-\hat{x}_0P_i(k)\right) e^{-k_0} \hat{e}_k. \\
 M_i\vartriangleright\hat{e}_k&=& i\left(\epsilon^{ijk} P_{j}(k) \hat{x}_{k}\right) e^{-k_{0}}  \hat{e}_k
\eea
Let us introduce the derivative operators on momentum space as follows
\be
\nabla^{0}\equiv \frac{\partial}{\partial k_{0}}- k_{i}\frac{\partial}{\partial k_{i}},\quad \nabla^{0}\equiv \frac{\partial}{\partial k_{i}}.
\ee
It can be checked that these derivatives implement the right multiplication on the group, that is
$$ \nabla^{\mu} \he_{k} = i \hat{x}^{\mu} \he_{k}$$
and the generators of Lorentz transformation can be written
\be\label{16}
  N_i\vartriangleright\hat{e}_k=
 e^{-k_0}\left( P_0(k)\nabla_{i}-P_i(k)\nabla_{0}\right)  \hat{e}_k, \quad
 M_i\vartriangleright\hat{e}_k = e^{-k_{0}}\left(\epsilon^{ijl} P_{j}(k) \nabla_{l}\right)   \hat{e}_k
\ee
One sees that the Lorentz  transformations acting on  $k$ are deformed and non linear, indeed
\begin{equation}\label{17}
  [M_i, k_j] = i\, \epsilon_{ijk} k_k, \quad [M_i, k_0] =0
\end{equation}
\begin{equation}\label{18}
   \left[N_{i}, {k}_{j}\right] = i\,  \delta_{ij}
 \left( {1\over 2} \left(
 1 -e^{-2{k_0}}
\right) + {{\mathbf{k}^2}\over 2}  \right) - i\, k_{i}k_{j} ,
\quad
  \left[N_{i},k_0\right] = i\, k_{i}.
\end{equation}
which are just the defining relations of $\kappa$-Poincar\'e algebra in the bicrossproduct basis \cite{kappaM1}.
However it can be easily checked that it follows from (\ref{17}), (\ref{18}) that the variables $P_\mu$ in (\ref{8}) transform as components of Lorentz vector, while $P_4$ is a Lorentz scalar.
\be
[N_{i},P_{j}(k)]= iP_{0}(k), \quad [N_{i},P_{0}(k)]= iP_{i}(k),\quad  [N_{i},P_{4}(k)]= 0.
\ee
In the field theory applications it is convenient therefore to label plane waves by these variables, instead of $k$.

The last technical point to be discussed here is the construction of differential calculus. To this end we should construct the infinitesimal translations $d\hat x^\mu$, and investigate the algebra they have with positions. It is a fundamental requirement that this algebra should be Lorentz covariant, so let us start with the way the differentials transform under Lorentz transformations. One should naturally require that for boosts
\begin{equation}\label{19}
   N_i \triangleright d\hx^\mu = d\left(N_i \triangleright \hx^\mu\right)
\end{equation}
and similarly for rotations. It follows from (\ref{2}) that the differentials transform in exactly the same way as positions. Now it is tempting to make use of the fact that we have already proved that the algebra (\ref{1}) is Lorentz covariant and take
$$
[\hx^0, d\hx^i] =id\hx^i
$$
with all other commutators vanishing. This does not work however since it turns out that the commutator $[\hx^i, d\hx^i] =0$ is not covariant under boost. The way out of this puzzle is to introduce one more differential $d\hx^4$, invariant under all Lorentz transformations, $N_i \triangleright d\hx^4 =M_i \triangleright d\hx^4=0$ \cite{5dcalc1}. It turns out that the algebra of positions and differentials takes the form
\begin{equation}\label{20}
   [\hx^\mu, d\hx^A] = (x^\mu)^A{}_B\, d\hx^B, \quad A,B=0,\ldots4
\end{equation}
where $(x^\mu)^A{}_B$ is the 5D matrix representation of positions (\ref{5}). Obviously (\ref{20}) is covariant, which can be checked by direct calculation.

Knowing what the algebra of differentials $d\hx^A$ is we can define the differential calculus by taking
\begin{equation}\label{21}
    d f(\hx) = d\hx^A\, \hP_A f(\hx)
\end{equation}
It can be checked by direct be tedious calculation that
\begin{equation}\label{22}
    \hP_A\, \he_k = P_A(k)\, \he_k
\end{equation}
where  $P(k)$ is given by (\ref{8}).

Let us discuss consequences of eq.\ (\ref{22})  more carefully. Notice first that the eigenvalues of derivatives $\hP_A$ can be decomposed into $P_\mu$ forming Lorentz vector and $P_4$ being Lorentz scalar. Thus, as in the standard case, $\hP^\mu\, \hP_\mu$ is a Lorentz invariant wave operator, which can be made equal $-m^2$, as usual. Then it follows that the group element $\he_k$ satisfies the standard field equation of massive (or massless) scalar field, so that it deserves the name plane wave. In the next section we will make use of this fact, defining the scalar field through (Fourier) decomposition into plane waves.

It has been argued in the recent paper \cite{AmelinoCamelia:2007vj} that since $\hP_0$ does not vanish on time ($\hx^0$) independent functions, it is not a generator of time translation and $P_0$ cannot be strictly speaking called energy (cf.\ (\ref{8})). However $P_0$ has the virtue that it forms, together with $P_i$, a Lorentz vector and this simple Lorentz property is, in our view, a good argument to choose it.
Moreover  one can devise a notion of time for which $P_{0}$ is the translation generator.

What is true is the fact that the notion of time translation depends on the choice of differential calculus. The question at hand is then which one leads to the most convenient notion of time and time translation and correspondingly which notion of energy is the preferred one.
The authors of \cite{Agostini:2006nc}, \cite{AmelinoCamelia:2007vj} seem to prefer quite arbitrarily the choice of time based on a specific ordering of plane wave.
But this is an arbitrary choice. Suppose for instance that we order the plane wave by putting the time on the left we have the identity
$\he_{k}= e^{ik_{0}\tilde{x}^{0}} e^{ik_{i}\hx^{i}}$ where the new ``time'' is $\tilde{x}^{0}= \hx^{0} +k_{i}\hx^{i}$.
and clearly a function independent of time $\hx^{0}$ is not independent of time $\tilde{x}^{0}$.

What we want to argue is that the choice of time and the corresponding energy should not be based on an arbitrary choice but govern by the symmetries and the dynamics of the theory under study.
As we have seen, the Lorentz symmetry favor the choice of $P_{0}$ as a time translation generator but even more than that the dynamics also favor a choice of a covariant time translation.
As we will see the canonical generator of time translation is also given by the covariant time translation.

This concludes our brief discussion of technical background. We refer the reader interested in more details to the paper \cite{Freidel:2007hk}. Let us now turn to more physical questions concerning construction of scalar field on $\kappa$-Minkowski space and its properties.

\section{Field theory on $\kappa$-Minkowski space}

Let us now now turn to construction of dynamical fields living on $\kappa$-Minkowski space. Since this space is non-commutative we must be careful about ordering.
Given a ("time to the right ordered" -- which means that in all expressions $x^0$ is moved to the right) field $\hat{\phi} = :\phi(\hat{x}):$\footnote{The space of fields is the space of functions that can be expressed as Fourier transform, i.e.\ the basis of this space is provided by plane waves $\he_k$.} we can define the translation invariant  integral to be
\be
\int_{\mathbb{R}^{4}} \hat{\phi}\equiv \int \mathrm{d}^{4}x\, \phi({x}).
\ee
where $\mathbb{R}^{4}$ denotes $\kappa$-Minkowski, while the integral on the right hand side is taken over the standard Minkowski space. This integral is the unique integral invariant under translation
\be
\int_{\mathbb{R}^{4}} \hat{k}_{\mu}\vartriangleright  \hat{\phi}=0.
\ee
where $\hat{k}_{\mu}\vartriangleright \he_k = k_\mu\, \he_k$.

It should be noticed that this integral is not cyclic since
\be
\int_{\mathbb{R}^{4}} \he_{k}\he_{p} = \delta(k_{0}+p_{0})\delta^{3}(\mathbf{k}+e^{-k_{0}}\mathbf{p})
= e^{3k_{0}} \delta(p_{0}+k_{0})\delta^{3}(\mathbf{p}+e^{-p_{0}}\mathbf{k})
= e^{3k_{0}}\int_{\mathbb{R}^{4}} \he_{p}\he_{k}
\ee
However it satisfy the exchange property
\be
\int_{\mathbb{R}^{4}}\he_{k}^{\dagger} \he_{p} =\int_{\mathbb{R}^{4}}\he_{p}^{\dagger} \he_{k}
\ee
and this property extends to functions, which can be expressed as Fourier integrals. In the formula above
\begin{equation}\label{23}
   \he_{k}^{\dagger} = \he_{S(k)}, \quad S(k_0) = - k_0, \quad S(k_i) = - k_i\, e^{k_0}
\end{equation}
is the (deformed) conjugation.

Using this integral we can define the Fourier coefficients  and the inverse Fourier transform
 to be
\be \label{24}
\tilde{\phi}(k)= \int_{\mathbb{R}^{4}} \he_{S(k)} \hat{\phi}, \quad\quad  \hat{\phi} =\int_{B}\mathrm{d}\mu(k) \,\, \he_{k} \tilde{\phi}(k)
\ee
where $B$ denotes the Borel group $\mathrm{d}\mu(k)=\frac{e^{3k_{0}}}{(2\pi)^{4}}\mathrm{d}{k_{0}} \mathrm{d}^{3}{\mathbf{k}}$ is the left invariant measure on it,
$\mathrm{d}\mu(pk)=\mathrm{d}\mu(k)$.

The conjugation of plane waves extends directly to conjugation of fields, to wit
\begin{equation}\label{25}
     \hf^\dagger(\hat{x})=\int \mathrm{d}\mu(k)  {\tilde\phi}^{*}(k)\,\he_{S(k)}
\end{equation}
where $*$ denotes the standard complex conjugation.

We will be interested in a free massive scalar theory, given by the Lorentz invariant  Lagrangian
\begin{equation}\label{26}
   \hat {\cal L} =\frac12\left[ (\hP_\mu \hf)^\dag \hP^\mu \hf + m^2\hf^\dag \hf\right]
\end{equation}
which leads to  the equation of motion
\be\label{27}
\hP_{\mu}\hP^\mu \hf + m^2 \hf=0
\ee
The action can be expressed in terms of Fourier modes as follows
\be\label{28}
 S= \int_{\mathbb{R}^{4}} \hat {\cal L} = \int \mathrm{d}\mu(k) \tilde{\phi}^{*}(k)\left(P^{\mu}P_{\mu}(k) + m^{2}\right)\tilde{\phi}(k)
\ee

Collecting together all the conditions that the on-shell state should satisfy, we get the following list

\begin{enumerate}
    \item The de Sitter space condition, following form the fact that points of Borel group belong to de Sitter space $P_AP^A =1$;
    \item The on shell condition following from (\ref{27}) $P_{\mu}P^\mu + m^2 =0$;
    \item The Borel group condition, cf.\ (\ref{8}), (\ref{9}), $P_0 + P_4 >0$.
\end{enumerate}

All these three conditions can be imposed by inserting the appropriate delta and Heaviside functions, as usual, see below. Let us now try to solve them algebraically. It follows from condition 1.\ and 2.\ that $P_4 = \pm \sqrt{1+m^2}$, and from condition 2.\ that $P_0 = \pm \sqrt{\mathbf{P}^2+m^2}\equiv \pm \omega_{\mathbf{P}}$. Imposing condition 3.\ we see that we have to do with three sectors, denoted as $+$, $-$, and $0$
\bea
\mbox{sector $+$} &:& P_{0} = +\omega_{\mathbf{P}},\quad P_{4} = +\sqrt{1+m^2} \nonumber\\
\mbox{sector $-$} &:& P_{0} = -\omega_{\mathbf{P}},\quad P_{4} = +\sqrt{1+m^2},\quad \bf{P}^{2}<1\label{29}\\
\mbox{sector $0$} &:& P_{0} = +\omega_{\mathbf{P}},\quad P_{4} = -\sqrt{1+m^2}, \quad \bf{P}^{2}>1\nonumber
\eea

These sectors are depicted on Fig.\ 1. Note that contrary to the standard case the momentum space is not simply connected, as it contains the trans-Planckian sector $0$\footnote{Recall that since $\kappa=1$, in sector $0$ momenta are larger than the scale $\kappa$, which is usually identified with the quantum gravity scale. For that reason we call these momenta trans-Planckian.}. Note also that as it is easy to see from (\ref{29}), the boundaries of sectors $-$ and $0$ are not Lorentz invariant. This can be seen also from Fig.\ 1, where Lorentz orbits are cross-sections of the de Sitter surface and the appropriate vertical planes; it follows that for sectors $-$ and $0$ these orbits necessarily cross the surface $P_0 +P_4=0$.

It should be stressed that when one takes as kinetic operator $(\hP_4-1)$ instead of $\hP_A\hP^A$, so that the on shell condition becomes $P_4-1 = M^2$, as it is done in the papers \cite{Agostini:2006nc}, \cite{AmelinoCamelia:2007uy}, \cite{AmelinoCamelia:2007vj}, the sector $0$ is missing, and the Lorentz invariance violation problem seems to be even more severe than in our case, see below.

Decomposing the field $\hf$ into modes described by three sectors (\ref{29}) we find
\begin{equation}\label{30}
    \hf= \int\frac{d^3P}{2\omega_{\mathbf{P}}
|P_4|}a_{+}(\mathbf{P})\he^{+}_{\mathbf{P}}+\int_{|\bf{P}|<1}\frac{d^3P}{2\omega_{\mathbf{P}}
|P_4|}a_{-}(\mathbf{\mathbf{P}})\he^{-}_{\mathbf{P}}
+\int_{|\bf{P}|>1}\frac{d^3P}{2\omega_{\mathbf{P}}
|P_4|}a_{0}(\mathbf{\mathbf{P}})\he^{0}_{\mathbf{P}}
\end{equation}
where $$\he^{\epsilon}_{\mathbf{P}}\equiv \he_{\left(P_{0}(\epsilon),\mathbf{P}(\epsilon),P_{4}(\epsilon)\right)},\quad \epsilon = +,-,0$$
and
$$P_{0}(+)=-P_{0}(-)= P_{0}(0)=\omega_{\mathbf{P}},\quad P_{i}(+)=-P_{i}(-)= P_{i}(0)={P}_{i},$$$$ P_{4}(+)= P_{4}(-)=-P_{4}(0)=\sqrt{1+m^{2}}$$
Notice that the momentum space is now not simply connected (cf.\ fig.\ 1) and thus although the last integral in (\ref{30}) looks like the first (with restricted integration range) in fact we are integrating over different parts of momentum manifold. It should be stressed again that had we chosen  $(\hP_4-1)$ as kinetic operator, the last term in expansion (\ref{30}) would be missing.

Having the field $\hf$ we can can compute the conjugate field $\hf^\dag$, by replacing $a$ with $a^*$, and the plane waves $\he_{\mathbf{P}}$ with $\he_{S(\mathbf{P})}$, where $S$ is the antipode defined by
$$
S(P)_{i}= -\frac{P_i}{P_{4}+ P_{0}},\quad
S(P)_{0}=-P_{0}+\frac{\mathbf{P}^2}{P_{0}+P_{4}}=-\frac{m^2+P_{0}P_{4}}{P_{0}+P_{4}} ,\quad S(P_{4})=P_{4}.
$$
It is important to note that the antipode exchanges the sectors $+$ with $-$ and maps $0$ onto itself and we denote by
$S^{\epsilon}_{P}$ the antipode restricted to these sectors. Thus, in the quantum field theory language, one could say that sectors $+$ and $-$ describe particles and antiparticles, respectively, while for sector $0$ particles are their own antiparticles.

In order to explicitly write down the conjugate field we will need to change variables $\mathbf{P}\to \mathbf{S^{\epsilon}_{P}}$.
Under this change of variable the measure transform as
\be\label{31}
\mathrm{d}^{3}S^{\epsilon}_{\mathbf{P}} = \mathrm{d}^{3}\mathbf{P} \mathrm{det}(\partial_{P_{i}}(\mathbf{S^{\epsilon}_{P}})_{j}))=
\frac{\mathrm{d}^{3} \mathbf{P}} {|P_{+}(\epsilon)|^{3}} \frac{\omega_{\mathbf{S^{\epsilon}_{P}}} }{\omega_{\mathbf{P}} }
\ee
with $P_{+}(\epsilon)=  P_{0}(\epsilon)+P_{4}(\epsilon)$.
Thus the conjugate field is given by
\bea\label{32}
\hf^\dagger&=&
\int\frac{d^3P}{2\omega_{\mathbf{P}} |P_4|}a_{+}^{*}(\mathbf{P})\he^{-}_{S^{+}_{\mathbf{P}}}+\int_{|P|<1}\frac{d^3P}{2\omega_{\mathbf{P}} |P_4|}{a_{-}^{*}(\mathbf{P})}\he^{+}_{S^{-}_{\mathbf{P}}}+\int_{|P|>1}\frac{d^3P}{2\omega_{\mathbf{P}} |P_4|}a_{0}^{*}(\mathbf{P})\he^{0}_{S^{0}_{\mathbf{P}}} \nonumber\\
&=&
\int\frac{d^3P}{2\omega_{\mathbf{P}} |P_4|}a_{-}^{\dagger}(\mathbf{P})\he^{+}_{\mathbf{P}}+
\int_{|P|<1}\frac{d^3P}{2\omega_{\mathbf{P}} |P_4|}{a_{+}^{\dagger}(\mathbf{P})}\he^{-}_{\mathbf{P}}+\int_{|P|>1}\frac{d^3P}{2\omega_{\mathbf{P}} |P_4|}a_{0}^{\dagger}(\mathbf{P})\he^{0}_{\mathbf{P}}
\eea
with
\bea
a_{-}^{\dagger}(\mathbf{P}) \equiv \frac{a_{-}^{*}(\mathbf{S^{+}_{P}})}{|P_{+}(+)|^{3}}, \quad
a_{+}^{\dagger}(\mathbf{P}) \equiv \frac{a_{+}^{*}(\mathbf{S^{-}_{P}})}{|P_{+}(-)|^{3}}, \quad
a_{0}^{\dagger}(\mathbf{P}) \equiv  \frac{a_{0}^{*}(\mathbf{S^{0}_{P}})}{|P_{+}(0)|^{3}}.
\eea
One now sees explicitly that positively ``charged'' particles are conjugate to negatively ``charged'' ones of bounded momenta $\mathbf{P}^{2}<1$, while the
trans-Planckian particles of type $0$  are self conjugate.

This concludes our discussion of on-shell fields. More details can be found in \cite{Freidel:2007hk}.

\section{The Noether charges}

Let us now turn to discussion of conserved charges associated with space-time symmetries of the theory. It should be stressed that only these charges really deserve the name of momenta and angular momenta, simply because they are conserved by construction. For this reason the Noether charges are expected to be related to observable quantities.

To construct the Noether charges one should consider the variation of the Lagrangian in the case when the variation of the field, denoted as $\delta\hf$ corresponds to a symmetry. In this case we know that the variation of the Lagrangian is to be, on-shell, given by a total derivative. Thus we must first decompose the variation of the Lagrangian into total derivative and a term proportional to field equations. In the case of our Lagrangian (\ref{26}) we have
 \be\label{33}
 \delta {\hat{\cal L}} = \hP_{A}
\left(\hat{\Pi}^A \delta \hf\right) + e^{\hk_{0}}\left((\hP_{\mu}\hP^\mu \hf + m^2
\hf)^\dagger \delta\hf \right) + \mathrm{h.c}
\ee
with canonical momenta being defined as follows
\bea
-\hat{\Pi}^{0}=\hat{\Pi}_0 &\equiv& \left(e^{-\hk_{0}}\hP_{0} \hf +
m^{2}\hf\right)^\dagger \label{34},\\
\hat{\Pi}^i=\hat{\Pi}_i &\equiv&
\left(\hP_{i}(1-e^{-\hk_{0}}\hP_{0})\hf\right)^\dagger,\\
\hat{\Pi}^4=\hat{\Pi}_4 &\equiv& \left(m^{2}\hf\right)^\dagger. \eea
It is worth noticing that although the zero component of field momentum (\ref{34}) looks unusual, by using the definition of conjugate derivatives
\be\label{35}
\hP_i^\dagger =-e^{-\hk_0}\hP_i, \quad \hP_0^\dagger=-\hP_0 +
\mathbf{\hP}^2e^{-\hk_0}, \hP_4^\dagger = \hP_4, \quad \left(e^{\hk_0}\right)^{\dagger}= e^{-\hk_0}
\ee
one can easily check that
\begin{equation}\label{36}
    \hat{\Pi}_0=\hP_4\, \hP_0 \hf^\dag
\end{equation}
which means that on-shell it differs from the standard time derivative of the field just by a constant multiplicative factor $\sqrt{1+m^2}$ (because on-shell $\hP_4\,  \hf^\dag = (\hP_4\,  \hf)^\dag = \sqrt{1+m^2}\,  \hf^\dag$).

Let us assume now that $\delta\phi=d_F\phi$, with $d_F$ being an appropriate differential, satisfying the Leibniz rule\footnote{In the case of translations $d_F = d\hx^A\, \hP_A$, ($A=0,\ldots,4$ since we are using the covariant differential calculus for translations, which happens to be five--dimensional, see \cite{Freidel:2007hk} for details and references); for Lorentz transformations $d_F =\omega^{\alpha\beta}\, L_{\alpha\beta}$, with  $L_{\alpha\beta}$ appropriate differential generators of these transformations, satisfying the standard algebra.}. Then we have
$$
\hP_{A} \left(\hat{\Pi}^A d_F \hf\right)+\hP_{A}^\dagger \left( (d_F
\hf)^\dagger\hat{\Pi}^{\dagger A}\right)-d_F\hat{\cal L}=0
$$
In the first term the differential of $\hf$ is placed to the right of the canonical momenta $\Pi$; and, of course these two terms, do not commute, since in general the transformation parameters do not commute with $\hx$. This problem can be easily solved by noticing that the differential $d_F$ satisfies Leibniz rule by definition, so that
\begin{equation}\label{37}
   \hP_{A} \left(d_F(\hat{\Pi}^A  \hf)-d_F\hat{\Pi}^A  \hf\right)+\hP_{A}^\dagger \left( (d_F
\hf)^\dagger\hat{\Pi}^{\dagger A}\right)-d_F\hat{\cal L}=0
\end{equation}

In order to calculate the charge associated with translations we specify $d_F = d\hx^A\, \hP_A$, use the covariance property $\hP_A\,d\hx^B=0$ that has been proved in \cite{Freidel:2007hk} and then disregard $d\hx$ to obtain the (on-shell) conservation equation
$$
\hP_{A} \left(\hP_B(\hat{\Pi}^A  \hf)-\hP_B\hat{\Pi}^A
\hf\right)+\hP_{A}^\dagger \left( \hP_B
\hf^\dagger\hat{\Pi}^{\dagger A}\right)-\hP_B\hat{\cal L}=0
$$
or
\begin{equation}\label{38}
    -\hP_{A} \left(\hP_B\hat{\Pi}^A \hf\right) +
\hP_{A}^\dagger \left( \hP_B\hf^\dagger\hat{\Pi}^{\dagger A}\right)
+\hP_B \left( \hP_{A}(\hat{\Pi}^A  \hf)- \hat{\cal L}\right)=0
\end{equation}
This equation can be reexpressed in the form
$$
\hP_{A} T^{A}{}_{B}=0
$$
where the components of the energy momentum tensor have the following form
\bea
T^{0}{}_{B}&=&-\hP_B\hat{\Pi}^0 \hf -\hP_{B}\hf^{\dagger} \Pi^{0\dagger} \\
T^{i}{}_{B}&=&-\hP_B\hat{\Pi}^i \hf - e^{{-k_{0}}}(\hP_{B}\hf^{\dagger} \Pi^{i \dagger})
+e^{{-k_{0}}}\hP^{i}(\hP_{B}\hf^{\dagger} \Pi^{0 \dagger})  \\
T^{4}{}_{B}&=&-\hP_B\hat{\Pi}^4 \hf +\hP_{B}\hf^{\dagger}\Pi^{4\dagger}
= 0
\eea
where in the last equation we use the explicit expression of $\Pi^{4}$.
Because of the last equality above, we just have the $4$-dimensional conservation equations
\begin{equation}\label{39}
    \hat{\partial}_\mu{T}^\mu_B=0
\end{equation}
in spite of the fact that the calculus we were using was five-dimensional. It can be shown that this property holds also in the case of interacting (and not just free) fields.

Now it is pretty straightforward, although quite tedious, to find the explicit form of conserved charges for translations. They are given by the formula
$$
\mathcal{P}_B=\int_{\mathbb{R}^{3}}  T_B^0 = -\int_{\mathbb{R}^{3} } (\hP_B\hat{\Pi}^0\hf+\hP_B\hf^\dagger\hat{\Pi}^{\dagger 0}).
$$
and read
\bea
{\cal{P}}_{0} &=&
\int_{\epsilon} \frac{d^3\mathbf{P} }{ 2\omega_{\mathbf{P}} |P_4|}
 \left( N_{+}(\mathbf{P}) + N_{-}(\mathbf{P}) - N_{0}(\mathbf{P})\right) \omega_{\mathbf{P}} \\
{\cal{P}}_{4} &=&
- \int_{\epsilon} \frac{d^3\mathbf{P} }{ 2\omega_{\mathbf{P}} }
 \left( N_{+}(\mathbf{P}) - N_{-}(\mathbf{P}) + N_{0}(\mathbf{P})\right) \\
 {\cal{P}}_{i} &=&
 \int_{\epsilon} \frac{d^3\mathbf{P} }{ 2\omega_{\mathbf{P}}|P_4| }
 \left( N_{+}(\mathbf{P}) - N_{-}(\mathbf{P}) + N_{0}(\mathbf{P})\right)P_{i}
 \eea
 where $N$'s are constructed from Fourier coefficients so that they corresponds to particle number operators in quantum theory. Explicitly
 $$N_{{\epsilon}}(\mathbf{P})=  a_{-\epsilon}^{\dagger}(\mathbf{P})a_{-\epsilon}^{}(\mathbf{S^{\epsilon}_P}). $$

 Let us pause here to discuss the meaning of these equations. First of all since for each mode we have the energy $\omega_{\mathbf{P}} \equiv \sqrt{m^2 + \mathbf{P}^2}$ and the momentum $\mathbf{P}$, we see that (in the quantum field theory language) for a single particle state the standard dispersion relation $P_0^2 - \mathbf{P}^2 =m^2$ holds. Thus, in agreement with earlier analyzes (for discussion see \cite{Kowalski-Glikman:2004qa} and references therein) there is no deformation of dispersion relation and, in particular no energy dependence of the speed of light. In fact in the present formulation most of the traces of deformation will be detectable only at the interacting theory level (e.g.\ modification of the conservation law in the vertex.)

 Second it seems that we are having a problem since the particle of type $0$ have negative energy.
However  the number of particle of type $0$ (again using the quantum field theory terminology) is also conserved because it can be expressed as a combination of conserved charges  $-2{\cal{N}}_{0}= \sqrt{1+m^{2}}{\cal Q}+ {\cal P}_{4}$ where
$${\cal Q}=-\int (\Pi^{0}\hf -\hf^{\dagger}\Pi^{0\dagger})= \int \frac{d^3\mathbf{P} }{ 2\omega_{\mathbf{P}} |P_4|}
\left(N_{{+}}(\mathbf{P})-N_{{-}}(\mathbf{P})-N_{{0}}(\mathbf{P})\right)$$
is the $U(1)$ charge.
Therefore in spite of the negative energy of sector $0$ modes
no instability can occur.

The charges associated with Lorentz transformations can be calculated in a similar way. The rotational charges have the standard form
\begin{equation}\label{40}
   {\cal{M}}_{ij} =\frac1i\,
 \sum_{\epsilon} \int_{\epsilon} \frac{d^3\mathbf{P} }{ 2\omega_{\mathbf{P}} |P_4|}
\alpha(\epsilon)   |{P_{+}(\epsilon)}|^{3} P_{[j}(\epsilon)  \left( \frac\partial{\partial P^{i]} }\, a_{-\epsilon}^{\dagger}(\mathbf{P})\right) a_{-\epsilon}^{\dagger *}(\mathbf{P})
\end{equation}
where $,\quad \alpha(+)=+1,\quad \alpha({-})=-1,\quad\alpha({0})=-1
.$

In the case of charges associated with Lorentz symmetry the situation is more complex, since in addition to the standard bulk term
\begin{equation}\label{41}
  {\cal N}_{i}^{bulk} =  -\frac1i\, \sum_\epsilon\int_{\epsilon} \frac{d^3\mathbf{P} }{ 2\omega_{\mathbf{P}} |P_4|} \alpha(\epsilon) \, P_+(\epsilon)\left[ \frac{P_i(\epsilon)}{P_{0}(\epsilon)}\,N_{\mathbf{P}}(\epsilon)+
  \omega_{\mathbf{P}}\, |{P_{+}(\epsilon)}|^{3}  \, \left(\frac{\partial}{\partial P_i} \, a_{-\epsilon}^{\dagger}(\mathbf{P})\right) a_{-\epsilon}^{\dagger *}(\mathbf{P})\right]
\end{equation}
 they acquire boundary terms, corresponding to the boundary of sectors $-$ and $0$ discussed earlier, to wit
 \begin{equation}\label{42}
 {\cal N}_{i}^{boundary}= \frac1i\, \int_{|\mathbf{P}|=1} \frac{d\Omega }{ 2 |P_4|}\, P_i \left( N_-(\mathbf{P}) - N_0(\mathbf{P})\right)
\end{equation}
where $d \Omega$ is the measure on the (momentum) unit sphere.

Note that while the contribution of $+$ sector to the boost charge is standard, it contains the nonstandard boundary term for both $-$ and $0$ sectors. These contributions would cancel if we glue together the boundary of the $-$ and $0$ sectors, i.e., if we assume that the particle disappearing from sector $-$, as a result of applying boost (we must apply a finite boost not an infinitesimal one to achieve this) reappears in sector $0$, and vice versa.

One sees that by gluing boundaries of sectors $-$ and $0$ in momentum space it is possible to save Lorentz symmetry. It should be stressed that such procedure is simply impossible in the models of scalar field theory on $\kappa$-Minkowski space considered in the series of papers by Amelino-Camelia et.\ al.\ \cite{Agostini:2006nc}, \cite{AmelinoCamelia:2007uy}, \cite{AmelinoCamelia:2007vj}, because in the case of the model considered there the sector $0$ is missing whatsoever, and the Lorentz symmetry is hopelessly lost\footnote{Of course identifying the generators of a symmetry is a mathematical statement and does not guarantee that the corresponding operationally defined quantities can be constructed. However, vice versa, if even the mathematically speaking the symmetry is missing there is no hope to construct its operational counterpart. Notice also that the effect of breaking Lorentz symmetry in sector $-$ has been already noticed in the one of the first papers on Doubly Special Relativity \cite{Bruno:2001mw}.}. This fact indicates that the model considered in these papers is not very interesting, as long as we have no reason to believe that in nature we have to do with an explicit breaking o Lorentz symmetry at Planck scale (for example exhibiting itself in the form of disappearance of antiparticles boosted to Planck energy.)

\section{Conclusions}

In this contributions we presented some themes described in our recent paper \cite{Freidel:2007hk}. Let us conclude with presenting a couple of the most important questions that are still left unanswered.

\begin{enumerate}
\item The issue of Lorentz symmetry. Infinitesimally the theory is perfectly Lorentz symmetric: it cannot see the boundaries of the region in momentum space. However this symmetry is at least endangered in the case of finite boosts. It is extremely interesting to investigate this problem further. What happens to the particles that disappear? If they really do what about energy/momentum conservation? If the effect of sector $-$/sector $0$ transmutation is real, what would be its observable signatures?
\item The interacting fields. The construction presented here and in \cite{Freidel:2007hk} should in principle hold in the case of interacting theories as well. However as a result of the fact that the integral over $\kappa$-Minkowski space is not cyclic it is not completely clear if an interacting theory, $\phi^3$ or $\phi^4$ say, possesses all the symmetries of the free one, considered here.
\end{enumerate}

\section*{Acknowledgment}

For JKG  this research was supported in part by the 2007-2010 research project N202 081 32/1844.

\end{document}